%% file: main.tex
\documentclass{article}
\usepackage[T1]{fontenc}
\usepackage[utf8]{inputenc}
\usepackage{lmodern}
\DeclareUnicodeCharacter{2212}{\ensuremath{-}}
\DeclareUnicodeCharacter{00D7}{\ensuremath{\times}}

\usepackage{graphicx}
\usepackage{amsmath}
\usepackage{amssymb}
\usepackage{array}
\usepackage{calc}
\usepackage{booktabs}
\usepackage{longtable}
\usepackage{caption}
\usepackage{xcolor}
\usepackage{authblk}
\usepackage[a4paper, total={170mm, 257mm}]{geometry}
\usepackage{natbib}
\usepackage{hyperref}
\title{Austerity, Gender, and Multidimensional Deprivation Among Older Europeans: Evidence from SHARE\thanks{Conference draft, prepared for the 30th Anniversary Conference of the Forum for Macroeconomics and Macroeconomic Policies (FMM), 22--24 October 2026. Please do not cite or distribute without permission from the author.}}
\author[1]{Nita Handastya}
\affil[1]{Department of Economics and Management ``Marco Fanno'', University of Padua}
\date{October 2026}

\begin{document}
\maketitle

\begin{center}
\fcolorbox{orange!70!black}{yellow!12}{%
  \parbox{0.92\textwidth}{\small\itshape\color{orange!40!black} Conference draft, prepared for the 30th Anniversary Conference of the Forum for Macroeconomics and Macroeconomic Policies (FMM), Institut für Makroökonomie und Konjunkturforschung, 22--24 October 2026. This version incorporates the completed empirical analysis (Sections 5 and 6) and the full estimation appendix; a small number of robustness checks noted in Sections 4.3 and 6.3 are ongoing. Please do not cite or distribute without permission from the author.}%
}
\end{center}

\begin{abstract}
This paper studies whether fiscal consolidation is associated with deeper multidimensional deprivation among older women than among older men, and through which channel. Using six waves of SHARE across eleven European countries, it builds a multidimensional deprivation index spanning functional health, economic insecurity, and social exclusion, and relates it to the cyclically adjusted primary balance (CAPB) at the country-wave level in a two-way fixed-effects model. Consolidation raises deprivation on average with a null average gender gap, which is later found to conceal regional heterogeneity where the gender gap is positive and significant in Southern and Eastern Europe, and absent in Continental/Nordic Europe. This regional pattern holds under individual clustering but does not survive clustering at the country level. Further, the economic insecurity channel the literature would predict runs in the opposite direction, with men’s measured insecurity rising more steeply than women’s; this is the only dimension whose interaction survives country-level inference. Social exclusion exhibits a larger gender effect, concentrated in the Eastern and Southern Europe, but it is significant only under individual clustering. The gendered cost of austerity among older Europeans is therefore neither uniform across welfare regimes nor routed through income, though the evidence for its social channel is inconclusive.

\end{abstract}

\section{Introduction}

The 2010s wave of fiscal consolidation across Europe has been well studied. The literature have examined its effects on output, employment, and, more recently, income inequality \citep{fatas2018, konstantinou2024}. The body of work, however, has been slower to ask who within the household and across the life course absorbs the adjustment. Public expenditure cuts do not fall evenly on individuals, instead they are amplified by existing patterns of employment, care responsibility, and entitlement, which are themselves gendered. A growing number of studies on gender disparity argues that austerity is not gender-neutral \citep{rubery2015, kushi2018, perugini2019}, although they focus overwhelmingly on working-age labour markets and household income. Little is known about how consolidation is experienced by older women and men. This is an important distinction to be taken into consideration, as their exposure runs through different and more concentrated channels. the first is pension entitlement which is accumulated (or not) over a working life, the second is reliance on public health and long-term care, and the third is unpaid care work, which is common at an age when people both give and receive care.

These channels matter because old age is where life-course disadvantage accumulates. For most women, their lifetime earnings and pension entitlements might reflect decades of interrupted participation, often caused by part-time or unpaid caregiving \citep{elson2016}. This disadvantage is exposed when public pensions, healthcare, and long-term care are cut, as often happens during consolidation. Cash-based measures of hardship capture the income side of this exposure but they say little about functional health, service access, or social participation. These dimensions become salient once people leave the labour market and depend more on public provision. Therefore, a purely income-based reading of austerity's welfare cost is likely to underestimate its gendered incidence among older cohorts.

This paper aims to address the gap with the use of a multidimensional deprivation framework applied to a longitudinal individual level data on living condition of the older European. Following the counting approach to multidimensional poverty measurement \citep{alkire2011}, it builds an equal-weight index with three domains: functional health, economic insecurity, and social exclusion. The index, built with the the Survey of Health, Ageing and Retirement in Europe (SHARE) data, covers individuals aged 50 and over in eleven European countries. Fiscal consolidation is measured by the cyclically adjusted primary balance (CAPB) at the country-wave level, matched to respondents by country and interview year. This combination allows a question that sits between the intersection of gendered effect of macroeconomic policy and quality-of-life outcomes. Does fiscal tightening widen the gender gap in a broader conception of later-life wellbeing? And if so, through which channels?

Two questions guide the analysis. First, is fiscal consolidation associated with higher multidimensional deprivation among older Europeans, and is the association stronger for women than for men? Second, where a gender gap exists, is it channeled through the economic, health, or social-exclusion dimension? Does it vary with welfare-regime type and consolidation intensity across Europe?

The paper aims to make two contributions. First, it extends the gendered-austerity literature, which has concentrated on employment and household income, to a multidimensional later-life outcome. It uses the Alkire-Foster counting approach, which is well established in development economics but rarely applied to fiscal policy questions in high-income Europe. Second, it combines harmonised individual-level panel data with a macro fiscal indicator across three welfare-regime groups. This allows the gender gap to be decomposed by mechanism and by institutional context, instead of being treated as one continent-wide average.

The remainder of the paper is organised as follows. Section 2 reviews the literature on the macroeconomics of fiscal consolidation, its gendered incidence, and multidimensional deprivation among older people. It also sets out the conceptual framework. Section 3 describes the data, the index, the country grouping, and the fiscal variable. Section 4 sets out the empirical model, the estimation strategy, and the limitations. Section 5 reports the results. Section 6 discusses them, returns to the limitations, and concludes.

\section{State of the Art}

\subsection{Fiscal consolidation: contested effects and a contested measure}

The literature on fiscal consolidation has moved a long way from the `expansionary fiscal contraction' hypothesis. Using a narrative measure of discretionary adjustments, \citet{guajardo2014} show that consolidations are contractionary. The earlier expansionary findings, they argue, reflected the endogeneity of cyclically adjusted balances to the cycle. \citet{fatas2018} extend the multiplier methodology of \citet{blanchard2013} to a longer horizon. They find that consolidations after the global financial crisis produced hysteresis effects large enough to be self-defeating in debt-to-GDP terms. This is consistent with Post-Keynesian scepticism of consolidation-led adjustment, and with the view that demand shocks can leave a permanent mark on potential output. \citet{konstantinou2024} adds the distributive dimension. Using dose-response methods, the study shows that consolidation episodes raise income inequality.

The composition of the adjustment matters as much as its size. \citet{heimberger2020} finds that consolidation raises income inequality. The effect is larger when the package is large, relies on spending cuts rather than tax increases, and follows a financial crisis in a low-growth period. These conditions describe the post-2010 adjustments in Southern Europe, but they also fit Eastern Europe, where post-crisis consolidation came on top of the earlier retrenchment of the transition years. Cuts of this kind fall on the public health, care, and social services through which this paper's channels run. \citet{karanikolos2013} document rising unmet medical need and deteriorating health across crisis-hit Europe. The deterioration was sharpest in the South, where austerity met already thin social protection. The composition of consolidation, and not the fiscal stance alone, is therefore what plausibly links a country-level shock to gendered later-life deprivation.

\subsection{The gendered incidence of austerity}

The literature has long argued that the incidence of fiscal adjustment should not be interpreted as aggregate accounts. \citet{elson2016} frames this as a problem of the gender-blind macroeconomic frameworks used in EU fiscal governance. In these frameworks cuts to public services are treated as costless, and as a result care work is assumed to be reabsorbed by the household, overwhelmingly by women, at no cost. \citet{rubery2015} documents how post-2008 austerity in Europe disrupted the earlier trend toward gender equality in employment. It combined public-sector job and pay cuts with cuts to the care services that had supported women's labour-force participation. The job and pay cuts fall mostly on women, who are the majority of public-sector employees.

This gendered employment cost is not uniform across welfare regimes. \citet{kushi2018} show that austerity's negative employment effects on women are largest where care provision was already residual and family-reliant before consolidation. \citet{perugini2019} extend the argument to wages. They find that consolidation in Europe widened gender wage gaps net of composition effects. This is direct evidence that the burden of adjustment is not shared equally between women and men, even where aggregate wage inequality does not move. \citet{bargawi2017} model the counterfactual instead. In a structuralist macroeconomic simulation, a gender-aware expansionary alternative generates substantially more employment for women than either continued austerity or a gender-neutral expansion. The status quo path thus carries a gendered opportunity cost, even where policymakers do not frame it that way.

A parallel literature in social epidemiology reaches similar conclusions by a different route. \citet{beckfield2018} analyse gender health equity as a function of social-investment policy, active labour-market programmes, and long-term care across Europe where they find that the health returns to social-investment spending are not gender-neutral. Retrenchment in these domains therefore falls unevenly on men and women. It is one route by which the gendered cost of consolidation is carried.

It is worth asking why a macro-level fiscal shock should reach individuals through gendered channels at all. The underlying mechanism typically focuses on the household-economics dimension. \citet{becker1965} established that unpaid domestic and care work is itself a productive use of time, not a residual left over once paid work is accounted for. Households allocate this time according to the relative returns available to each member in the labour market, and when public provision of care or health services contracts, someone has to absorb the resulting workload. The collective household models of \citet{chiappori1992}, and their extension by \citet{lundberg1996}, predict that who absorbs it is not random. It depends on each partner's bargaining position, which is a function of earnings, pension entitlements, and outside options built up earlier in the life course.

This is why the gender gap should be expected to widen with age, not stay constant over the life course. The bargaining position that determines who bears a fiscal shock in later life is the cumulative product of decades of unequal labour-market attachment. \citet{kleven2019} document the origin of that inequality in Danish administrative data. The arrival of children creates a long-run earnings penalty for women of around 20 percent, driven by reduced hours, participation, and wages. The penalty compounds over the working life rather than fading. A gender pension gap in later life is thus the accumulated record of decades of child and caregiving penalties. A fiscal shock arriving at age 50 or later can therefore be read, in part, as the deferred cost of decisions and constraints faced decades earlier.

None of this literature follows the cohort concerned into old age. The employment and wage channels they study focuses mainly on the individual's working life during their productive age. For people already aged 50 or over when consolidation hits, the more relevant channel is pension income, which is fixed by past contribution histories that are largely gendered. The effect is amplified by cuts to sectors such as public health and long-term care, and they affect older women more severely because of their informal-care obligations. Many are simultaneously care recipients and unpaid carers for their families. This means that as women accumulated disadvantages throughout their productive ages, it still continues to their retirement. 

\subsection{Multidimensional deprivation and the life course}

Measuring the welfare cost of austerity through income poverty alone risks missing the dimensions of hardship most relevant to an older population. The counting approach of \citet{alkire2011} offers an alternative. Individuals are identified as deprived by counting the indicators, across several domains, in which they fall below a threshold. As such, a single monetary cutoff is not needed. The approach was developed for poverty measurement in low- and middle-income countries, but it adapts well to ageing in high-income Europe. For people aged 50 and over, functional health, service access, and social participation are no less central to material wellbeing than income.

SHARE is well suited to this exercise. It is a harmonised, cross-nationally comparable panel on health, socioeconomic status, and social and family networks for people aged 50 and over \citep{borschsupan2013}. It has already been used to document the health consequences of austerity-era retrenchment. Using SHARE, \citet{borboudaki2024} find that preventive healthcare use fell and reported lack of access to care rose across the austerity period and the deterioration was sharpest in Southern Europe. The study treats health service use as the outcome, although it does not disaggregate by gender or combine health with economic and social-participation indicators in a single measure. The richness of SHARE allows a more granular analysis, which this paper aims to exploit.

\subsection{Welfare-regime context and the analytical framework}

Since the seminal work of \citet{espingandersen1990}, there have been many attempts to revisit the classification of European welfare states. This paper does not enter the debate on the most suitable grouping. It groups countries by fiscal-consolidation intensity and welfare-regime type. The groups are South (Italy, Spain, Greece), East (Estonia, Poland, the Czech Republic, Slovenia), and Continental/Nordic (Germany, France, Sweden, Denmark).

\citet{ferrera1996} provides the analytical basis for the Southern group. They identifies Italy, Spain, Greece, and Portugal as a distinct `Southern' welfare model. It is characterised by fragmented income maintenance, strong reliance on the family as a substitute for thin public care provision, and clientelistic elements in welfare administration. The group shares a Troika-era experience of externally conditioned, front-loaded consolidation after 2010. This came on top of a familialist regime in which informal, largely female, family care substitutes for public provision.

\citet{fenger2007} provides the corresponding basis for the Eastern group. They cluster post-socialist welfare states along dimensions of decommodification and defamilisation, and identifies a shared post-communist consolidation trajectory running from the 1990s reforms onward. This trajectory is structurally earlier and less externally conditioned than the Southern group. It is no less consequential for the coverage of the public pensions and care services on which the current cohort of older Europeans relies. Slovenia does not place cleanly in his typology.  However, other literature such as \citet{aidukaite2011}, puts it in a broader `post-communist regime' classification, and this paper follows that view.

The Continental/Nordic group is the comparison baseline. These are universalist, social-insurance-based welfare states \citep{espingandersen1990} with moderate consolidation intensity over the sample period. The size of any gender gap in the other two groups is benchmarked against them. The cluster analysis of defamilisation regimes by \citet{bambra2007} supports this placement. It puts the Continental and Nordic countries at the opposite end of the family-care-substitution spectrum from the Southern group.

Figure 1 visualises the conceptual framework linking the literature to the empirical strategy. The mechanism this paper relies on assumes that fiscal consolidation reaches the composite deprivation index through three channels: labour-market and pension, care and social reproduction, and health and long-term care. Each channel is proxied by one of the three sub-indices developed based on \citet{alkire2011} approach, and is also aggregated as the composite multidimensional deprivation index. Gender moderates the channels through the $CAPB \times Female$ interaction. The same fiscal shock is thus transmitted with different intensity, depending on the accumulated life-course position of the person it reaches. Welfare-regime grouping and wave fixed effects adds further layer of analysis to further study the effect of policy approach on the transmission of austerity to the gender gap in the wellbeing outcomes. 

\begin{figure}[htbp]
\centering
\includegraphics[width=0.92\textwidth]{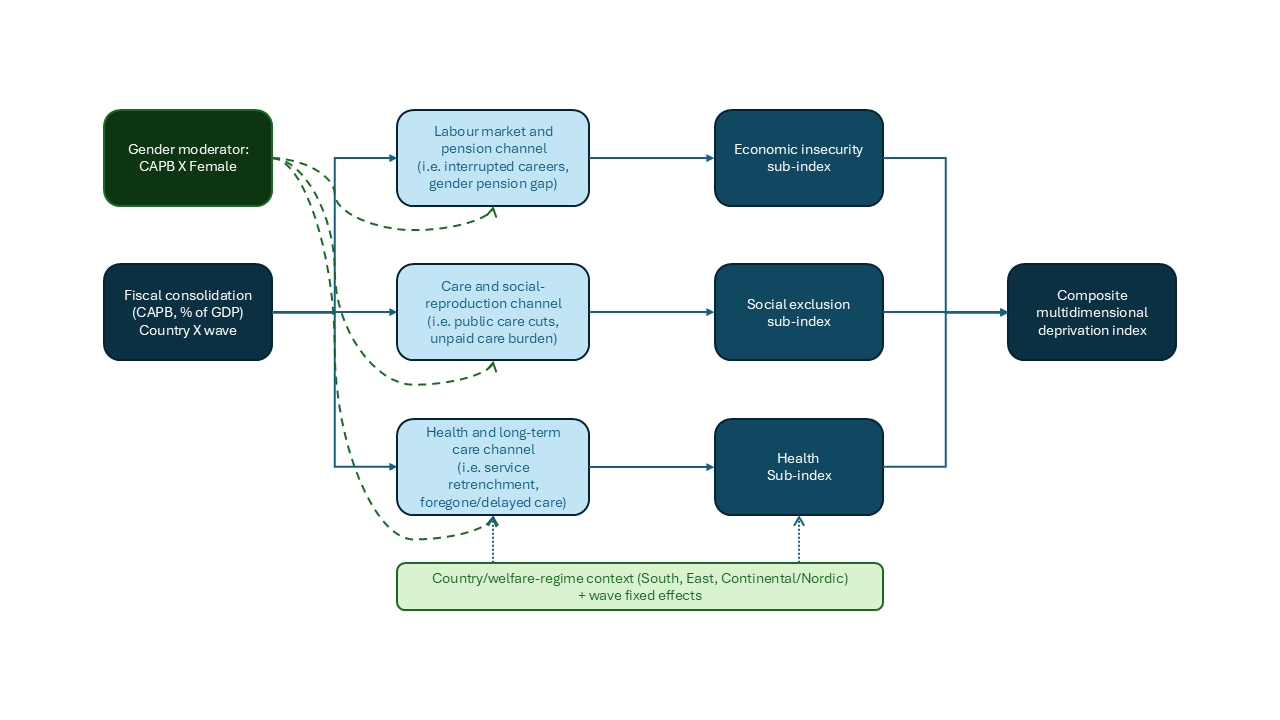}
\caption{Conceptual pathways linking fiscal consolidation to multidimensional deprivation, by gender}
\end{figure}

\section{Data}

\subsection{Sample and data source}

This study uses data from the Survey of Health, Ageing and Retirement in Europe (SHARE) data. SHARE is a longitudinal, multidisciplinary, cross-national panel of individuals aged 50 and older which collects detailed information on health, socioeconomic status, and social participation across a growing set of European countries \citep{borschsupan2013}.

The analysis uses six of SHARE's nine waves: Wave 1 and Waves 5 to 9. Wave 3 (SHARELIFE) is excluded because it holds only retrospective life-history information. Waves 2 and 4 are dropped because the missing indicator is crucial for the development of the economic deprivation sub-index (i.e. foregone medical or dental care due to cost). Although the fieldwork of wave 9 was partly affected by the COVID-19 pandemic, this study retains its use. Wave fixed effects absorb any average wave-level shift, and a specification without Wave 9 is reported in the appendix.

The sample covers eleven countries in three groups. South: Italy, Spain, and Greece. East: Estonia, Poland, the Czech Republic, and Slovenia. Continental/Nordic: Germany, France, Sweden, and Denmark. The reasoning for the grouping has been explained in the previous section. 

The final analytic sample has 181,861 person-wave observations on 77,905 individuals in eleven countries and six waves. The unit of observation is the person-wave. Standard errors are clustered at the individual level in the main analysis. The appendix examines sensitivity to this choice.

\subsection{The multidimensional deprivation index}

The multidimensional deprivation index used in this study follows the  counting approach of \citet{alkire2011}. A composite index is built as an equal-weighted average of the sub-indices, namely: health, economic, and social sub-indices, scaled to range between 0 to 1. Following is the detailed explanation of each dimension. 

\emph{Health sub-index.} It is comprised of three binary indicators drawn from the current-status health module: (i) poor self-rated health, coded 1 if the respondent rates their health as fair or poor; (ii) a chronic-conditions indicator, coded 1 if the respondent reports two or more of the standard list of chronic conditions used throughout the SHARE health module; and (iii) a functional-limitation indicator, coded 1 if the respondent reports at least one limitation on either the ADL or IADL activity. 

\emph{Economic sub-index.} It is comprised of three binary indicators: (i) difficulty in making ends meet is coded 1 when the response is affirmative, (ii) household income in the bottom income quartile, in order for the indicator to reflects relative economic position given local cost-of-living and income distributions, and (iii) Foregone medical or dental care due to cost is coded 1 when the response is affirmative.

\emph{Social sub-index.} It is comprised of two binary indicators: (i) (i) living alone, taken from the individual level coverscreen module in which the variable indicating household member is coded as 1. This study disregards different possible living arrangements (house/apartment/farm house, elderly living complex, nursing home) ; and (ii) non-participation in social or leisure activities in the past twelve months, based on the standard activities in the activities module, coded 1 if the respondent selects ‘none of these’ across the full list of activities, voluntary or charity work, training or education courses, sport or social clubs, religious organisations, and political or community organisations.

The composite deprivation index is the simple average of the three sub-indices, ranging from 0 (no deprivation on any indicator) to 1 (deprivation on every indicator across all three domains). Lower figure indicates fewer deprivation and vice versa. 

\subsection{The fiscal variable}

Fiscal consolidation is measured by the level of the cyclically adjusted primary balance (CAPB), expressed as a percentage of GDP. It is sourced from the AMECO database and matched to SHARE respondents by country and interview year. A higher CAPB indicates a more austere fiscal stance. In the analytical sample, CAPB ranges from -4.9 to 9.9\% percent of GDP, with a mean of 0.23 and standard deviation of 2.75 (Table 1). 

The measure has a known weakness. \citet{heimberger2017} argue that the potential-output estimates behind cyclically adjusted balances are procyclical and path-dependent. They tend to understate the stance of fiscal policy exactly when consolidation is most severe. The critique targets foundations shared by the IMF, the European Commission, and the majority of the consolidation literature. However, it is not, in itself, a reason to abandon CAPB. CAPB remains the most widely available cross-country, cross-wave fiscal indicator that can be matched to the SHARE panel. As such, CAPB coefficient in this study should be interpreted as a plausible lower bound on the intensity of consolidation and not a precise estimate.

\subsection{Descriptive overview}

Table 1 reports descriptive statistics for the composite deprivation index, the three sub-indices, and their component indicators, alongside the fiscal variable and age, disaggregated by sex. Women score higher than men on the aggregate multidimensional deprivation index where the values are respectively 0.294 and 0.244. The same gap is present in each of the three sub-index: health, economic, and social. Women in the sample are almost twice more likely to live alone compared to men and slightly more likely to experience difficulty in making ends meet. The average age between the two groups are very similar, with an overall sample at 68.52 years old. Among the three sub-index, the most prevalent one is health followed by economic and social. When it comes to the austere average fiscal environment over their observed waves, women are less affected compared to men.  

\small
\begin{longtable}[]{@{}
  >{\raggedright\arraybackslash}p{(\columnwidth - 6\tabcolsep) * \real{0.5000}}
  >{\raggedright\arraybackslash}p{(\columnwidth - 6\tabcolsep) * \real{0.1667}}
  >{\raggedright\arraybackslash}p{(\columnwidth - 6\tabcolsep) * \real{0.1667}}
  >{\raggedright\arraybackslash}p{(\columnwidth - 6\tabcolsep) * \real{0.1667}}@{}}
\caption{Sample characteristics by sex}\\
\toprule
\textbf{} & \textbf{Male} & \textbf{Female} & \textbf{Total} \\
\midrule
\endhead
Multidimensional deprivation index (0--1) & 0.244 & 0.294 & 0.272 \\
Health sub-index & 0.345 & 0.390 & 0.370 \\
Economic insecurity sub-index & 0.204 & 0.248 & 0.229 \\
Social exclusion sub-index & 0.184 & 0.244 & 0.217 \\
Lives alone & 0.148 & 0.282 & 0.224 \\
Difficulty making ends meet & 0.351 & 0.396 & 0.376 \\
Age at interview & 68.62 & 68.44 & 68.52 \\
CAPB (\% of GDP) & 0.286 & 0.188 & 0.231 \\
\textbf{Observations} & \textbf{79,381} & \textbf{102,480} & \textbf{181,861} \\
\bottomrule
\multicolumn{4}{@{}p{\dimexpr\columnwidth-2\tabcolsep\relax}@{}}{\footnotesize\emph{Means. The full indicator list is in the appendix.}}
\end{longtable}

\section{Methodology}

\subsection{Empirical model}

To examine the association between fiscal consolidation and multidimensional deprivation, and whether it differs by gender, the paper estimates a linear model by OLS. The dependent variable is the composite deprivation index and, separately, each of its three sub-indices. Following is the model:

\begin{equation}
    Deprivation_{ict} = \alpha + \beta_{1} CAPB_{ct} + \beta_{2} Female_{i} + \beta_{3} CAPB_{ct} \times Female_{i} + \gamma X_{ict} + \delta_{c} + \tau_{t} + \varepsilon_{ict}
\end{equation}

Where $\text{Deprivation}_{ict}$ is the composite (or sub-index) score for individual $i$ in country $c$ at wave $t$. $\text{CAPB}_{ct}$ is the cyclically adjusted primary balance for country $c$ at wave $t$. $Female_{i}$ is a gender indicator. $X_{ict}$ is a vector of individual covariates. $\delta_c$ and $\tau_t$ are country and wave fixed effects. $\varepsilon_{ict}$ is an error term. The coefficient of interest of this paper is $\beta_{3}$, on the $CAPB \times Female$ interaction. It captures whether consolidation is associated with a larger increase in deprivation for women than for men, net of country and wave effects and of the covariates in $X$.

The composite index is continuous on {[}0, 1{]}, so a linear specification is a good fit. Coefficients can be interpreted directly as changes in the deprivation score. Where the outcome is the binary headcount (deprived in at least one third of weighted indicators), the corresponding linear probability model is reported in the appendix.

\emph{Controls (X).} Age, age squared, education (ISCED-97, four categories), and employment status. Household size, partnership status, and marital status are excluded. `Lives alone' is one of the two social-exclusion indicators, so controlling for household composition would net out part of the outcome. 

\subsection{Estimation strategy and inference}

The analysis has four stages. The first estimates the baseline model on the pooled sample, building up from a bivariate CAPB specification to the full interaction model. It then repeats the full model for the composite index and each sub-index. The second stage estimates the full model separately for the South, East, and Continental/Nordic subsamples, without a three-way interaction. It does so for the composite and for each sub-index. The third stage adds the bottom-income-quartile indicator to the health sub-index model. It tests whether a consolidation-driven gender gap in health survives once income position is netted out. The fourth stage checks robustness. It uses a one-year lag, the year-on-year change, a pre-2020 window, a specification without Wave 9, and a binary headcount outcome.

Standard errors are clustered at the individual level in the main analysis. CAPB varies only at the country-wave level. Individual clustering may therefore understate the standard errors on a regressor that is constant within larger groups \citep{moulton1990}. The main specification is re-estimated with clustering at the country-wave and country level. Eleven countries are too few for reliable country-level asymptotics \citep{cameron2008}. The interaction terms are therefore also tested with a wild cluster bootstrap with Webb weights and 9,999 replications. The bootstrap clusters by country for the pooled models and by country-wave for the regional models, which have only three or four countries each.

\subsection{Limitations}

Several limitations should be noted. First, CAPB is a country-year aggregate used to proxy an individual-level fiscal shock. This is a standard limitation in the consolidation literature. Second, CAPB rests on potential-output estimates that \citet{heimberger2017} show to be procyclical and path-dependent. As Section 2.1 explains, the CAPB coefficient is read as a plausible lower bound on the intensity of consolidation. The measure is retained for want of a comparable cross-country, cross-wave alternative.

Third, the sample has no observations between Wave 1 (2004/05) and Wave 5 (2013). For countries observed in Wave 1, variation in CAPB comes from comparing the pre-crisis wave with the later waves, and from changes within Waves 5 to 9. The build-up of consolidation after 2008 is not observed directly.

Fourth, household size, partnership, and marital status are excluded from the controls (Section 4.1) to avoid tautology with the social exclusion sub-index. Unobserved selection into these characteristics may be correlated with both CAPB exposure and deprivation. The social exclusion sub-index carries the paper's central gender-gap estimate, so this bears directly on the headline result. Section 6.3 returns to it. Employment status is retained despite being a plausible bad control in the sense of \citet{angrist2009}. It may respond to the same fiscal shocks that the model evaluates. Its inclusion is a stated limitation, not one resolved by omitting or instrumenting it.

Fifth, and most fundamentally, the identification strategy is associational, not causal. CAPB is not randomly assigned across countries and waves, and no instrument isolates exogenous variation in fiscal stance. The two-way fixed-effects specification absorbs time-invariant country characteristics and common wave shocks. It does not absorb country-specific trends that are correlated with both consolidation and deprivation. The paper is therefore a contribution to measurement and to the gendered-incidence debate. It does not offer a causal estimate of the effect of austerity on deprivation.

\section{Results}

\subsection{The baseline association and the pooled gender gap}

Table 2 presents the baseline specification. CAPB is positive and significant in every column. A one percentage point tighter primary balance is associated with an increase of 0.0027 to 0.0030 in the composite index. The estimate is stable when the female indicator, the full controls, and the interaction term are added. Women score about 0.029 higher than men on the composite index, conditional on controls and fixed effects. The female coefficient is significant in every specification. A one standard deviation (2.75 percentage points) tighter balance raises the index by about 0.0075. This is 2.8 percent of the sample mean (0.272) and 3.6 percent of its standard deviation (0.207).

\small
\begin{longtable}[]{@{}
  >{\raggedright\arraybackslash}p{(\columnwidth - 8\tabcolsep) * \real{0.3830}}
  >{\raggedright\arraybackslash}p{(\columnwidth - 8\tabcolsep) * \real{0.1543}}
  >{\raggedright\arraybackslash}p{(\columnwidth - 8\tabcolsep) * \real{0.1543}}
  >{\raggedright\arraybackslash}p{(\columnwidth - 8\tabcolsep) * \real{0.1543}}
  >{\raggedright\arraybackslash}p{(\columnwidth - 8\tabcolsep) * \real{0.1543}}@{}}
\caption{Fiscal consolidation, gender, and multidimensional deprivation}\\
\toprule
\textbf{} & \textbf{(1)} & \textbf{(2)} & \textbf{(3)} & \textbf{(4)} \\
\midrule
\endhead
CAPB (\% of GDP) & 0.0028*** & 0.0028*** & 0.0030*** & 0.0027*** \\
& (0.0002) & (0.0002) & (0.0002) & (0.0003) \\
Female & -- & 0.0467*** & 0.0287*** & 0.0286*** \\
& & (0.0014) & (0.0012) & (0.0012) \\
CAPB × Female & -- & -- & -- & 0.0005 \\
& & & & (0.0003) \\
Controls / country \& wave FE & FE only & FE only & Full & Full \\
\textbf{Observations} & 181,861 & 181,861 & 181,861 & 181,861 \\
R-squared & 0.073 & 0.085 & 0.275 & 0.275 \\
\bottomrule
\multicolumn{5}{@{}p{\dimexpr\columnwidth-2\tabcolsep\relax}@{}}{\footnotesize\emph{Standard errors in parentheses, clustered at the individual. Controls: age, age squared, education, employment status. * p\textless0.10, ** p\textless0.05, *** p\textless0.01. Appendix Table A2 reports all coefficients for column 4.}}\\
\end{longtable}

The interaction term enters in column 4. Its coefficient is small and insignificant. It stays insignificant under clustering at the individual, country-wave, or country level. It also stays insignificant under a wild cluster bootstrap that allows for the small number of country clusters (t(10) = 0.755, p = 0.439; 95 percent confidence set {[}$-$0.0014, 0.0017{]}). Appendix Table A6 gives the full comparison. The marginal effect of CAPB is slightly larger for women (0.0032) than for men (0.0027), and the confidence intervals overlap. Figure 2 plots predicted deprivation by CAPB and sex. The lines are close to parallel across the observed range of the fiscal variable.

\begin{figure}[htbp]
\centering
\includegraphics[width=0.75\textwidth]{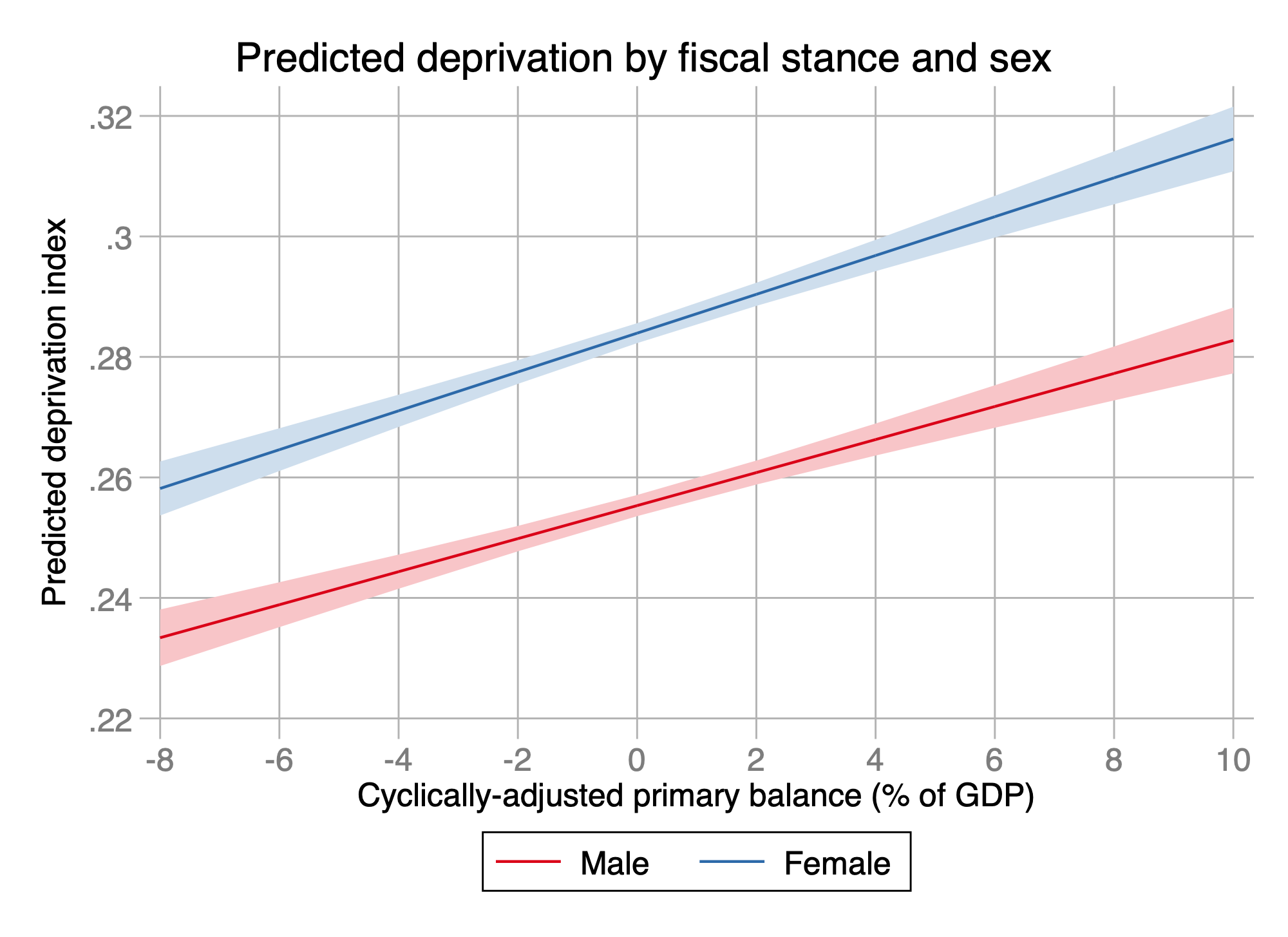}
\caption{Predicted composite deprivation index by CAPB and sex (95\% CI)}
\end{figure}

\subsection{Dimensions and the gender gap}

Table 3 re-estimates the full model for the composite index and each sub-index. CAPB is positive and significant in every sub-index except social exclusion. Its effect is largest for economic insecurity (0.0057). On average, consolidation therefore registers most directly as economic hardship. Its effect on social exclusion is small and insignificant. The female coefficient is largest in the social exclusion sub-index (0.0487).

The interactions point in different directions. In the economic insecurity model the interaction is negative and significant ($-$0.0021). Men's economic insecurity rises more steeply with consolidation than women's. In the social exclusion model the interaction is positive and significant (0.0024). Women's social exclusion rises with consolidation, while men's does not. The health interaction is weakly positive (0.0011, p = 0.055). Figure 3 plots the interactions by dimension.

\small
\begin{longtable}[]{@{}
  >{\raggedright\arraybackslash}p{(\columnwidth - 8\tabcolsep) * \real{0.3830}}
  >{\raggedright\arraybackslash}p{(\columnwidth - 8\tabcolsep) * \real{0.1543}}
  >{\raggedright\arraybackslash}p{(\columnwidth - 8\tabcolsep) * \real{0.1543}}
  >{\raggedright\arraybackslash}p{(\columnwidth - 8\tabcolsep) * \real{0.1543}}
  >{\raggedright\arraybackslash}p{(\columnwidth - 8\tabcolsep) * \real{0.1543}}@{}}
\caption{Sub-indices and the gender gap}\\
\toprule
\textbf{} & \textbf{Composite} & \textbf{Health} & \textbf{Economic} & \textbf{Social} \\
\midrule
\endhead
CAPB (\% of GDP) & 0.0027*** & 0.0019*** & 0.0057*** & 0.0006 \\
& (0.0003) & (0.0005) & (0.0004) & (0.0004) \\
Female & 0.0286*** & 0.0199*** & 0.0173*** & 0.0487*** \\
& (0.0012) & (0.0021) & (0.0015) & (0.0019) \\
\textbf{CAPB × Female} & \textbf{0.0005} & \textbf{0.0011*} & \textbf{−0.0021***} & \textbf{0.0024***} \\
& (0.0003) & (0.0006) & (0.0004) & (0.0005) \\
Observations & 181,861 & 181,861 & 181,861 & 181,861 \\
R-squared & 0.275 & 0.204 & 0.178 & 0.159 \\
\bottomrule
\multicolumn{5}{@{}p{\dimexpr\columnwidth-2\tabcolsep\relax}@{}}{\footnotesize\emph{Standard errors in parentheses, clustered at the individual. Full controls and country/wave FE in all columns. * p\textless0.10, ** p\textless0.05, *** p\textless0.01. Appendix Table A2 reports all coefficients. Appendix Table A6 (panel A) reports the same interactions under country-wave and country clustering and a wild cluster bootstrap.}}\\
\end{longtable}

The economic interaction is the reverse of the mechanism anticipated in Section 2.2. The household-economics literature suggests an income and pension channel running against women, but the estimate runs the other way. The social exclusion interaction instead points toward a care and social-reproduction channel.

Stricter inference agrees with this result. CAPB varies at the country-wave level, so individual clustering may be too permissive. Under country-wave clustering the social interaction is no longer significant (p = 0.135). That comparison uses 58 clusters, so the loss of significance is not a small-sample artefact. It is also insignificant under country clustering (p = 0.368) and under the wild cluster bootstrap (p = 0.525). The economic interaction survives all three (p = 0.029, 0.020, and 0.012). The health interaction is insignificant under all three. Appendix Table A6 (Panel A) reports the comparison. With only eleven countries, the tests must be read with caution, and therefore the social-exclusion result should be considered suggestive. The result is large when clustering is by individual and imprecise when clustering matches the level of variation in CAPB.

\begin{figure}[htbp]
\centering
\includegraphics[width=0.75\textwidth]{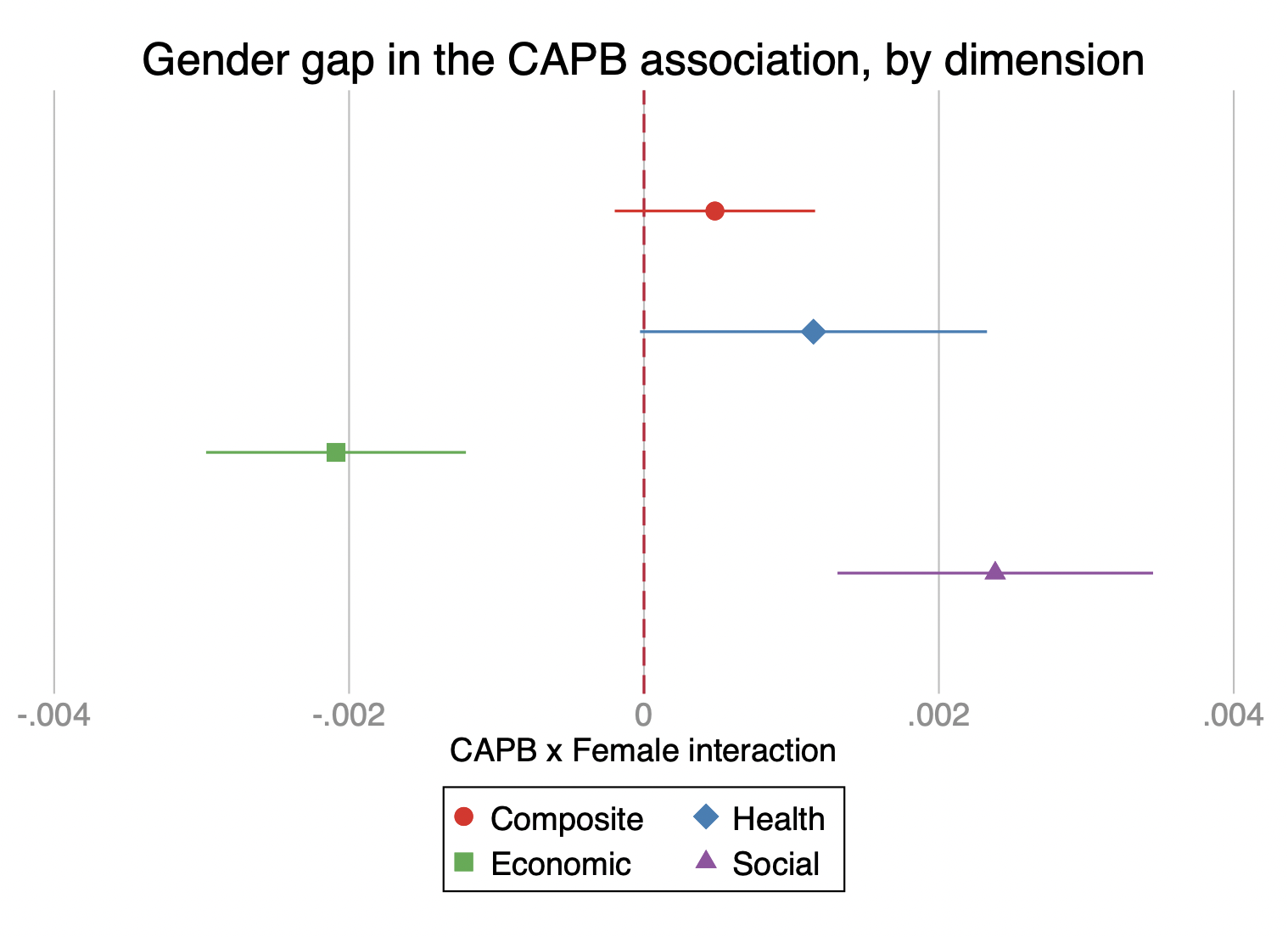}
\caption{$CAPB \times Female$ interaction by deprivation dimension (composite, health, economic, social; 95\% CI)}
\end{figure}

\subsection{Regional heterogeneity}

Table 4 re-estimates the full model by welfare-regime group, for the composite index and each sub-index. Appendix Table A3 reports the full estimates, including the main effects of CAPB and female by region.

\small
\begin{longtable}[]{@{}
  >{\raggedright\arraybackslash}p{(\columnwidth - 8\tabcolsep) * \real{0.2800}}
  >{\raggedright\arraybackslash}p{(\columnwidth - 8\tabcolsep) * \real{0.1800}}
  >{\raggedright\arraybackslash}p{(\columnwidth - 8\tabcolsep) * \real{0.1800}}
  >{\raggedright\arraybackslash}p{(\columnwidth - 8\tabcolsep) * \real{0.1800}}
  >{\raggedright\arraybackslash}p{(\columnwidth - 8\tabcolsep) * \real{0.1800}}@{}}
\caption{Gender gap by welfare-regime group and dimension (CAPB × Female interaction)}\\
\toprule
\textbf{Welfare-regime group} & \textbf{Composite} & \textbf{Health} & \textbf{Economic} & \textbf{Social} \\
\midrule
\endhead
South (n = 50,997) & 0.0022*** & 0.0005 & 0.0009 & 0.0053*** \\
& (0.0005) & (0.0008) & (0.0006) & (0.0008) \\
East (n = 59,472) & 0.0022*** & 0.0010 & 0.0016* & 0.0039*** \\
& (0.0006) & (0.0011) & (0.0008) & (0.0009) \\
Continental/Nordic (n = 71,392) & −0.0010 & 0.0005 & −0.0018* & −0.0016 \\
& (0.0008) & (0.0015) & (0.0010) & (0.0013) \\
\bottomrule
\multicolumn{5}{@{}p{\dimexpr\columnwidth-2\tabcolsep\relax}@{}}{\footnotesize\emph{Cell entries are the $CAPB \times Female$ coefficient (standard error), clustered at the individual, from separate full-model regressions of each outcome by region. * p\textless0.10, ** p\textless0.05, *** p\textless0.01. Appendix Table A3 reports all coefficients. Appendix Table A6 (panel B) reports country-wave clustering and a wild cluster bootstrap for the dimension estimates.}}\\
\end{longtable}

On the composite index, the interaction is positive and significant in the South and the East, while it is negative and insignificant in Continental/Nordic Europe ($-$0.0010, p = 0.25). This contrasts with the pooled null. The pooled average may hide different regional experiences. Consolidation is associated with a wider gender gap in the South and East, where welfare regimes are familialist and consolidation was historically most severe. In the universalist Continental/Nordic group the point estimate is slightly negative but not distinguishable from zero. The three regressions are run on separate subsamples, so the difference between regions is not itself tested. Figure 4 plots the interactions by region with the composite estimates using standard errors clustered at the individual level. The stricter tests in Appendix Table A6 (panel B) cover the dimension estimates only. The regional composite estimates have not been re-estimated at a stricter clustering level.

The dimension columns locate the gap within each region. In the South and East the gap runs through social exclusion (0.0053 and 0.0039, both p $<$ 0.01**, clustered at the individual). In Continental/Nordic Europe the social interaction is small, negative, and insignificant. The economic interaction there is negative ($-$0.0018). It is weakly significant under individual clustering (p = 0.077) and insignificant under country-wave clustering (p = 0.102) and the bootstrap (p = 0.106).

Stricter inference again weakens the social result. Under country-wave clustering the South social interaction remains significant (p = 0.034), but it does not survive the bootstrap (p = 0.103). The East social interaction is insignificant under the same clustering (p = 0.253; bootstrap p = 0.258). The regional pattern is therefore a difference in which dimension the gap appears to run through, and not only in the size of a common gap. That conclusion is firmest for the South.

\begin{figure}[htbp]
\centering
\includegraphics[width=0.75\textwidth]{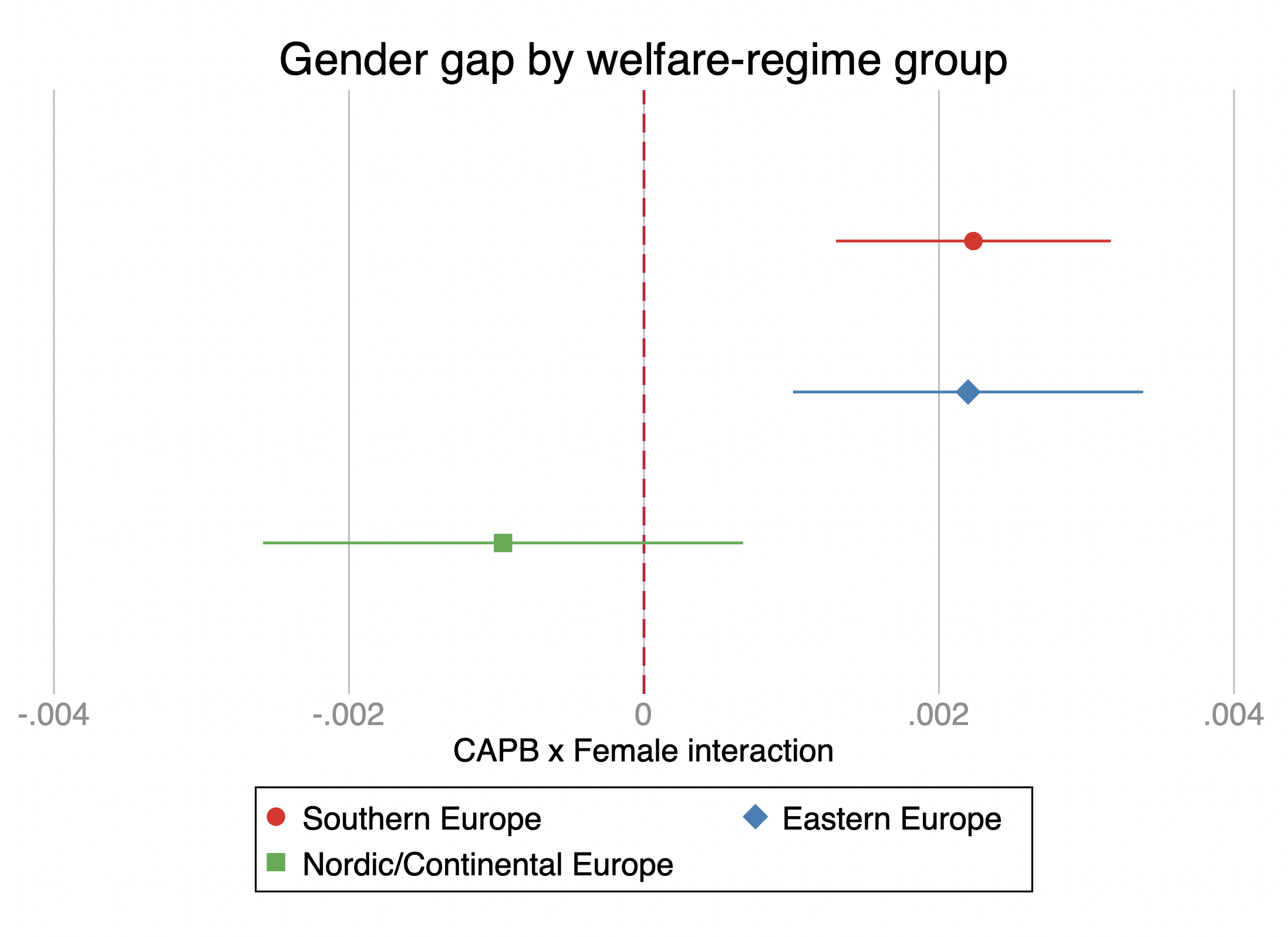}
\caption{$CAPB \times Female$ interaction by welfare-regime group (South, East, Continental/Nordic; 95\% CI)}
\end{figure}

\subsection{Robustness}

Three sets of checks support the headline results. All are reported in the appendix. The first varies the timing of the fiscal variable, the estimation window, and the outcome definition (Appendix Table A5). The CAPB coefficient remains positive and significant with a one-year lag, with the year-on-year change, in a pre-2020 window, without the pandemic-affected Wave 9, and with a binary headcount outcome. In the pre-2020 window it rises to 0.0045, possibly because the 2020 to 2022 fiscal expansion dilutes the pooled estimate. The pooled interaction stays close to zero and insignificant throughout.

The second adds the bottom-income-quartile indicator, and its interaction with CAPB, to the health sub-index model (Appendix Table A4). Income enters the economic sub-index, not the health sub-index, so the check is not circular. The health interaction does not shrink when income position is controlled; it strengthens slightly, from 0.0011 to 0.0012 and 0.0013. The interaction of income quartile with CAPB is small and insignificant, so consolidation is not associated with a steeper health gradient for the poorest. Because the health interaction is significant under individual clustering only, this rules out an income explanation instead of establishing a health channel.

The third varies the level of clustering (Appendix Table A6, panel A). Point estimates are identical across specifications, and only precision changes. The CAPB effect stays significant under country clustering (p = 0.021). The wild cluster bootstrap confirms that the pooled interaction is null. The two headline results, a robust average CAPB effect and a robust average gender-gap null, are not artefacts of a particular specification or clustering choice. Appendix Table A6 also reports the clustering comparison for the dimension and regional interactions, discussed in Sections 5.2 and 5.3, where the economic interaction survives at every level, while the social and regional interactions do not.

\section{Discussion and conclusion}

Fiscal consolidation is associated with higher multidimensional deprivation among older Europeans. The average gender gap in this association is a precisely estimated null. It survives every clustering choice and specification considered. The pooled null may hide regional differences. The estimated gap is positive in Southern and Eastern Europe and absent in Continental/Nordic Europe. The regional composite estimates use individual clustering only.

Decomposing the index by dimension locates the positive interaction in social exclusion and a negative interaction in economic insecurity. The negative sign means men’s measured economic insecurity rises more steeply with consolidation than women’s. The economic result holds under stricter inference. The social result does not. It is imprecise once standard errors are clustered at the level of variation in CAPB. The South social interaction holds under country-wave clustering but not under the bootstrap, and the East interaction does not hold. The evidence on mechanism is therefore suggestive. In Continental/Nordic Europe there is no evidence that consolidation widens the gender gap.

\subsection{The regional pattern}

Accounts of gendered austerity \citep{elson2016, rubery2015, kushi2018, perugini2019} predict a gender gap in the incidence of consolidation. The pooled null does not show it. The regional estimates are consistent with the prediction. The positive estimates appear where the conditions assumed by these accounts hold: familialist care provision and front-loaded, externally conditioned consolidation. In the universalist welfare states of the Continental/Nordic group \citep{espingandersen1990} the predicted gap does not appear. Welfare-regime type thus appears to matter for gendered deprivation under consolidation. The three regional models are estimated on separate subsamples, so the difference between regions is not formally tested. The pattern, however, does not extend to health. The health interaction is null in all three groups, so the gender-differentiated health returns to social-investment spending reported by \citet{beckfield2018} do not appear here. The gap is concentrated in the South and East, where social-investment retrenchment was most severe \citep{ferrera1996, fenger2007, bambra2007}.

A recent study of later-life poverty puts the regional result in context. \citet{hartull2024} find that older women are more exposed to objective income poverty than men, even with individual controls. Welfare-regime type is not a significant predictor of objective or subjective poverty in their models. Their finding describes a standing gender difference in the level of poverty. This paper finds that consolidation does not add to that difference on the economic dimension. The economic insecurity interaction is negative and robust, so men’s measured insecurity rises more steeply with consolidation. Considering both accounts, the two results locate the gender difference in the level of later-life poverty, but not in how consolidation moves it.

This also affects how consolidation should be read. CAPB records the size of an adjustment, not its incidence. Two countries can post the same improvement in the balance and impose very different costs, depending on what is cut and who depends on it. The pattern is consistent with the view that consolidation which cuts care and social services widens the gap between older women and men. The data cannot show this, because CAPB does not record what was cut. It is a hypothesis for future work. What a consolidation does to different groups depends on its composition, which the primary balance does not show.

\subsection{Channels affecting the gender gap: social exclusion and economic insecurity}

A regional gender gap raises the question of channel. The household-economics account is the natural starting point. It runs from \citet{becker1965} through \citet{chiappori1992} and \citet{lundberg1996} to the child-penalty evidence in \citet{kleven2019}. It predicts that a fiscal shock in later life falls disproportionately on women through the income and pension channel. The results run the other way. The economic insecurity interaction is negative: men's measured economic insecurity rises more steeply with consolidation than women's. This result holds under stricter inference, with p-values between 0.012 and 0.029. The positive interaction appears instead in social exclusion, which here means living alone and non-participation in social or leisure activity. That estimate is large under individual clustering and imprecise otherwise. It is suggestive, not established. If it holds, it challenges the view that income is what matters most for the multidimensional wellbeing of older Europeans.

The social exclusion index needs a closer reading. Women’s excess in this sub-index comes from living alone (0.282 for women, 0.148 for men). They are slightly less likely than men to report no social or leisure participation (0.205 against 0.220). The interaction is estimated on the two-item sub-index and is not split by indicator. The result therefore speaks to social exclusion, and not to lost participation. Living alone reflects widowhood and household formation, which the model leaves out of the controls. That exclusion is deliberate (Section 4.1) and is noted among the limitations in Section 4.3. Differential mortality between older women and men is an alternative interpretation.

There are two ways to read the social result. The first takes it at face value. Cuts to public care and social infrastructure may reach older women through reduced social contact more than through household income. Plausible routes are reduced public transport, closure of community and day-care centres, and thinning local social infrastructure. Loneliness is a known problem among older cohorts. Its drivers vary across European countries with economic conditions, welfare systems, and cultural expectations \citep{mercan2025association}. \citet{pagan2024dynamic} find that disability is a strong driver of social exclusion. Consolidation that cuts services for the most vulnerable can erode their capacity to engage socially. Independent evidence from SHARE points the same way. \citet{juma2024} build a multidimensional measure of social exclusion. They find that being a woman and living in an urban area raise the risk of exclusion. The risk is higher in Southern and Eastern Europe, and where income inequality is greater and trust in government lower. This is the population, outcome, and region in which this paper’s estimate is largest. It also agrees with \citet{elson2016}, who argues that austerity's costs are often invisible to income-based measurement. They are absorbed as unpaid time, instead of forgone earnings.

The second interpretation is more cautious about the economic sub-index. Its three components are: difficulty making ends meet, bottom income quartile, and foregone medical care due to cost. They are a coarse measure of economic hardship. They capture current hardship better than accumulated pension position, so the negative interaction focuses on hardship and not to lifetime entitlement. Thus, a gender gap that runs through pensions may not be captured by them. The caregiving route offers a more plausible explanation. \citet{savela2026} study material deprivation among older Europeans. They find that women who provide family care face a higher risk of deprivation, specifically material hardship and energy poverty. The caregiving role intensifies with age and weighs most on those already in a weaker position. A gap of this kind is partly economic and partly social, and the sub-indices used here do not separate the two. Both readings imply that an assessment of austerity's gendered cost that stops at income-based indicators may miss the channel to which these results point.

\subsection{Limitations revisited}

The results highlight limitations beyond those in Section 4.3.

The first concerns the fiscal measure. CAPB records the size of an adjustment and not its composition. It does not say what was cut or who relied on it. The reading that consolidation reached older women through cuts to care and social services is inferred from the pattern of results. Testing it would need spending data by function, so that the health, pension, and care components of the adjustment can be separated. CAPB captures the stance of policy but not its incidence, and incidence is where the gender gap arises.

The second concerns the period covered. The complete-case rule drops Waves 2 and 4, so the panel runs from 2004 to 2005 and then from 2013 to 2022. The years from 2006 to 2012 are absent. They contain the financial crisis and the front-loaded Troika programmes in the South. The sharpest consolidation episode in the sample period is therefore observed mainly in its aftermath. An available-case index would recover these two waves, at the cost of a two-item economic sub-index in the early period, and would return Portugal to the sample.

The third concerns the channel and the indicators that carry it. The study design is associational, so it establishes a pattern but does not identify the mechanism. The social-exclusion result is also imprecise under stricter clustering, as Section 5.2 shows. A cost that runs jointly through material and social routes, such as caregiving, cannot be attributed cleanly to one sub-index. It may surface on the social side. The social indicators raise further issues. Living alone and non-participation are self-reported, and what counts as engagement varies across countries with culture and welfare provision. Part of the regional pattern may reflect reporting differences. Household composition is also excluded from the controls, to avoid tautology with living alone. Selection into it may therefore be correlated with both consolidation exposure and deprivation. The result could partly reflect differences in widowhood between women and men across countries and waves, not consolidation.

The fourth concerns the regional results. The grouping rests on welfare-regime type and consolidation intensity, which are not measured directly. Welfare-regime typologies are contested, and the choice among them is not settled \citep{bambra2007, chybalski2021, saxonberg2021}. For gendered fiscal effects, typologies built on defamilisation and familialism fit best, because they capture how far care is provided by the family and not the state. The grouping would look different on that basis. This applies especially to Slovenia, which \citet{chybalski2021} also discusses, in a way similar to \citet{fenger2007}. A check that drops Slovenia, reported in the appendix, leaves the East result intact. Without Slovenia the East composite interaction is larger (0.0033, p = 0.002, against 0.0022). The East social interaction is also larger (0.0077, against 0.0039). Both use individual clustering. The direction of these changes is consistent with Slovenia being a poor fit for the East group. Inference is a further constraint here. Each region has only three or four countries, so clustering by country and the country-level bootstrap are not feasible. The regional composite estimates use clustering by individual, and the dimension estimates add clustering by country-wave and a bootstrap. The three regional models are also estimated on separate subsamples, so no test of the difference between regions is reported. The regional results should therefore be read as patterns in the data and not as precise estimates.

\subsection{Conclusion and directions for future research}

The paper asked whether fiscal consolidation widens the gender gap in a multidimensional measure of later-life deprivation among older Europeans, and through which channel. On average, consolidation is associated with higher deprivation for everyone, and the association is not gendered. Averaged across welfare regimes, this may hide a regional pattern. In Southern and Eastern Europe, where consolidation was most severe and care most familialised, women bear a larger estimated deprivation cost than men. The point estimates place that cost in social exclusion, not in the income and pension channel that the economics literature emphasises. This result is suggestive, because it is imprecise under stricter clustering. It is concentrated where welfare regimes lean on the family for care. On the economic dimension the finding is firmer, and it runs against the prediction: men’s measured insecurity rises more steeply with consolidation than women’s. Income-based measurement does not register the social cost, which may be one reason the gendered incidence of consolidation has been easy to overlook.

Two directions follow for future work. The first is measurement. Austerity should be read through the composition of spending, not through the primary balance alone. This would show which dimensions of spending carry the effect and trace the cost to the services that were cut. The outcome side needs the same treatment. Estimating the interaction separately for living alone and non-participation would show which indicator carries the social result. A larger country sample would also help, since eleven countries leave the regional estimates imprecise. The second is policy. EU fiscal governance and distributional impact assessment still rarely treat gender, ageing, and social infrastructure as intersecting axes of vulnerability. The results suggest that an income-focused view of austerity’s distributive cost may miss the channel, and the population, in which this paper finds the gendered cost most concentrated.

\bibliography{references}

\newpage
\input{appendix.tex}

\end{document}

%% file: appendix.tex
\appendix
\renewcommand{\thetable}{A\arabic{table}}
\setcounter{table}{0}

\section{Appendix}

The tables report the models in Section 5 in full. Standard errors are in parentheses and clustered at the individual level unless stated otherwise. * p\textless0.10, ** p\textless0.05, *** p\textless0.01. The last table reports the Slovenia check discussed in Section 6.

\small
\begin{longtable}[]{@{}
  >{\raggedright\arraybackslash}p{(\columnwidth - 6\tabcolsep) * \real{0.5000}}
  >{\raggedright\arraybackslash}p{(\columnwidth - 6\tabcolsep) * \real{0.1667}}
  >{\raggedright\arraybackslash}p{(\columnwidth - 6\tabcolsep) * \real{0.1667}}
  >{\raggedright\arraybackslash}p{(\columnwidth - 6\tabcolsep) * \real{0.1667}}@{}}
\caption{Sample characteristics by sex (full indicator list)}\\
\toprule
\textbf{} & \textbf{Male} & \textbf{Female} & \textbf{Total} \\
\midrule
\endhead
Composite deprivation index (0--1) & 0.244 & 0.294 & 0.272 \\
\quad Health sub-index & 0.345 & 0.390 & 0.370 \\
\quad \quad Poor self-rated health & 0.374 & 0.409 & 0.394 \\
\quad \quad Two or more chronic conditions & 0.490 & 0.529 & 0.512 \\
\quad \quad One or more ADL/IADL limitation & 0.172 & 0.232 & 0.206 \\
\quad Economic insecurity sub-index & 0.204 & 0.248 & 0.229 \\
\quad \quad Difficulty making ends meet & 0.351 & 0.396 & 0.376 \\
\quad \quad Bottom income quartile & 0.210 & 0.276 & 0.247 \\
\quad \quad Foregone care due to cost & 0.050 & 0.071 & 0.062 \\
\quad Social exclusion sub-index & 0.184 & 0.244 & 0.217 \\
\quad \quad Lives alone & 0.148 & 0.282 & 0.224 \\
\quad \quad No social or leisure participation & 0.220 & 0.205 & 0.211 \\
Age at interview & 68.62 & 68.44 & 68.52 \\
CAPB (\% of GDP) & 0.286 & 0.188 & 0.231 \\
\textbf{Observations} & \textbf{79,381} & \textbf{102,480} & \textbf{181,861} \\
\bottomrule
\multicolumn{4}{@{}p{\dimexpr\columnwidth-2\tabcolsep\relax}@{}}{\footnotesize\emph{Means. Indented rows are the 0/1 component indicators of each sub-index. Source: SHARE Waves 1, 5--9; author's calculations.}}
\end{longtable}

\normalsize

\footnotesize
\begin{longtable}[]{@{}
  >{\raggedright\arraybackslash}p{(\columnwidth - 8\tabcolsep) * \real{0.3600}}
  >{\raggedright\arraybackslash}p{(\columnwidth - 8\tabcolsep) * \real{0.1600}}
  >{\raggedright\arraybackslash}p{(\columnwidth - 8\tabcolsep) * \real{0.1600}}
  >{\raggedright\arraybackslash}p{(\columnwidth - 8\tabcolsep) * \real{0.1600}}
  >{\raggedright\arraybackslash}p{(\columnwidth - 8\tabcolsep) * \real{0.1600}}@{}}
\caption{Composite index and sub-indices: all coefficients}\\
\toprule
\textbf{} & \textbf{Composite} & \textbf{Health} & \textbf{Economic} & \textbf{Social} \\
\midrule
\endhead
CAPB (\% of GDP) & 0.0027*** & 0.0019*** & 0.0057*** & 0.0006 \\*
 & (0.0003) & (0.0005) & (0.0004) & (0.0004) \\
Female & 0.0286*** & 0.0199*** & 0.0173*** & 0.0487*** \\*
 & (0.0012) & (0.0021) & (0.0015) & (0.0019) \\
\textbf{CAPB $\times$ Female} & 0.0005 & 0.0011* & $-$0.0021*** & 0.0024*** \\*
 & (0.0003) & (0.0006) & (0.0004) & (0.0005) \\
Age & $-$0.0190*** & $-$0.0168*** & $-$0.0171*** & $-$0.0231*** \\*
 & (0.0007) & (0.0012) & (0.0009) & (0.0011) \\
Age squared (/100) & 0.0173*** & 0.0193*** & 0.0114*** & 0.0213*** \\*
 & (0.0005) & (0.0009) & (0.0006) & (0.0008) \\
Primary education & $-$0.0479*** & $-$0.0355*** & $-$0.0431*** & $-$0.0651*** \\*
 & (0.0031) & (0.0052) & (0.0039) & (0.0048) \\
Secondary education & $-$0.1083*** & $-$0.0947*** & $-$0.1120*** & $-$0.1183*** \\*
 & (0.0030) & (0.0050) & (0.0038) & (0.0046) \\
Tertiary education & $-$0.1636*** & $-$0.1542*** & $-$0.1831*** & $-$0.1536*** \\*
 & (0.0031) & (0.0052) & (0.0039) & (0.0048) \\
Employed or self-employed & $-$0.0442*** & $-$0.0877*** & $-$0.0577*** & 0.0127*** \\*
 & (0.0017) & (0.0029) & (0.0022) & (0.0025) \\
Other employment status & 0.0809*** & 0.0884*** & 0.1081*** & 0.0462*** \\*
 & (0.0020) & (0.0032) & (0.0025) & (0.0029) \\
Sweden & $-$0.0519*** & $-$0.1146*** & $-$0.0214*** & $-$0.0198*** \\*
 & (0.0027) & (0.0048) & (0.0032) & (0.0039) \\
Spain & $-$0.0123*** & $-$0.0824*** & $-$0.0122*** & 0.0577*** \\*
 & (0.0028) & (0.0048) & (0.0034) & (0.0040) \\
Italy & 0.0260*** & $-$0.0916*** & 0.0696*** & 0.1000*** \\*
 & (0.0029) & (0.0048) & (0.0036) & (0.0041) \\
France & $-$0.0047* & $-$0.0626*** & 0.0150*** & 0.0335*** \\*
 & (0.0028) & (0.0049) & (0.0035) & (0.0041) \\
Denmark & $-$0.0400*** & $-$0.0801*** & $-$0.0364*** & $-$0.0036 \\*
 & (0.0027) & (0.0050) & (0.0031) & (0.0041) \\
Greece & 0.0535*** & $-$0.0982*** & 0.1909*** & 0.0677*** \\*
 & (0.0029) & (0.0049) & (0.0037) & (0.0045) \\
Czech Republic & 0.0157*** & $-$0.0030 & 0.0265*** & 0.0237*** \\*
 & (0.0027) & (0.0048) & (0.0033) & (0.0039) \\
Poland & 0.0780*** & 0.0277*** & 0.1132*** & 0.0931*** \\*
 & (0.0030) & (0.0052) & (0.0039) & (0.0046) \\
Slovenia & 0.0226*** & $-$0.0479*** & 0.0949*** & 0.0208*** \\*
 & (0.0028) & (0.0049) & (0.0035) & (0.0040) \\
Estonia & 0.1023*** & 0.1133*** & 0.1438*** & 0.0498*** \\*
 & (0.0027) & (0.0046) & (0.0034) & (0.0038) \\
Wave 5 & $-$0.0511*** & 0.0147*** & 0.0093*** & $-$0.1774*** \\*
 & (0.0016) & (0.0028) & (0.0022) & (0.0027) \\
Wave 6 & $-$0.0603*** & 0.0017 & $-$0.0078*** & $-$0.1749*** \\*
 & (0.0017) & (0.0029) & (0.0022) & (0.0027) \\
Wave 7 & $-$0.0642*** & 0.0062 & $-$0.0236*** & $-$0.1753*** \\*
 & (0.0022) & (0.0038) & (0.0029) & (0.0034) \\
Wave 8 & $-$0.0514*** & 0.0041 & 0.0088*** & $-$0.1672*** \\*
 & (0.0018) & (0.0031) & (0.0024) & (0.0029) \\
Wave 9 & $-$0.0460*** & 0.0008 & 0.0007 & $-$0.1395*** \\*
 & (0.0018) & (0.0031) & (0.0024) & (0.0030) \\
Constant & 0.8638*** & 0.7196*** & 0.9019*** & 0.9698*** \\*
 & (0.0259) & (0.0444) & (0.0336) & (0.0398) \\
\midrule
Observations & 181{,}861 & 181{,}861 & 181{,}861 & 181{,}861 \\
Individuals & 77{,}905 & 77{,}905 & 77{,}905 & 77{,}905 \\
R-squared & 0.2749 & 0.2041 & 0.1775 & 0.1587 \\
\bottomrule
\multicolumn{5}{@{}p{\dimexpr\columnwidth-2\tabcolsep\relax}@{}}{\footnotesize\emph{Standard errors in parentheses, clustered at the individual. Reference categories: no education, retired, Germany, Wave 1. The composite column is the full model behind the baseline estimate. * p\textless0.10, ** p\textless0.05, *** p\textless0.01.}}
\end{longtable}

\normalsize

\footnotesize
\begin{longtable}[]{@{}
  >{\raggedright\arraybackslash}p{(\columnwidth - 8\tabcolsep) * \real{0.3600}}
  >{\raggedright\arraybackslash}p{(\columnwidth - 8\tabcolsep) * \real{0.1600}}
  >{\raggedright\arraybackslash}p{(\columnwidth - 8\tabcolsep) * \real{0.1600}}
  >{\raggedright\arraybackslash}p{(\columnwidth - 8\tabcolsep) * \real{0.1600}}
  >{\raggedright\arraybackslash}p{(\columnwidth - 8\tabcolsep) * \real{0.1600}}@{}}
\caption{Regional models: all coefficients}\\
\toprule
\textbf{} & \textbf{Composite} & \textbf{Health} & \textbf{Economic} & \textbf{Social} \\
\midrule
\endhead
\multicolumn{5}{@{}l}{\textit{Panel A. South}} \\
CAPB (\% of GDP) & 0.0057*** & 0.0054*** & 0.0085*** & 0.0032*** \\*
 & (0.0004) & (0.0008) & (0.0006) & (0.0008) \\
Female & 0.0220*** & 0.0435*** & $-$0.0239*** & 0.0463*** \\*
 & (0.0026) & (0.0044) & (0.0033) & (0.0041) \\
\textbf{CAPB $\times$ Female} & 0.0022*** & 0.0005 & 0.0009 & 0.0053*** \\*
 & (0.0005) & (0.0008) & (0.0006) & (0.0008) \\
Age & $-$0.0148*** & $-$0.0085*** & $-$0.0139*** & $-$0.0219*** \\*
 & (0.0013) & (0.0021) & (0.0017) & (0.0021) \\
Age squared (/100) & 0.0153*** & 0.0155*** & 0.0092*** & 0.0211*** \\*
 & (0.0009) & (0.0015) & (0.0012) & (0.0015) \\
Primary education & $-$0.0414*** & $-$0.0312*** & $-$0.0315*** & $-$0.0614*** \\*
 & (0.0037) & (0.0062) & (0.0047) & (0.0058) \\
Secondary education & $-$0.1110*** & $-$0.1002*** & $-$0.1107*** & $-$0.1222*** \\*
 & (0.0038) & (0.0062) & (0.0048) & (0.0058) \\
Tertiary education & $-$0.1549*** & $-$0.1330*** & $-$0.1756*** & $-$0.1561*** \\*
 & (0.0043) & (0.0071) & (0.0054) & (0.0067) \\
Employed or self-employed & $-$0.0037 & $-$0.0467*** & $-$0.0035 & 0.0390*** \\*
 & (0.0032) & (0.0052) & (0.0045) & (0.0051) \\
Other employment status & 0.0760*** & 0.0574*** & 0.1273*** & 0.0433*** \\*
 & (0.0030) & (0.0049) & (0.0039) & (0.0047) \\
Italy & 0.0299*** & $-$0.0164*** & 0.0715*** & 0.0346*** \\*
 & (0.0030) & (0.0049) & (0.0038) & (0.0044) \\
Greece & 0.0467*** & $-$0.0324*** & 0.1792*** & $-$0.0066 \\*
 & (0.0031) & (0.0053) & (0.0041) & (0.0051) \\
Wave 5 & $-$0.0796*** & $-$0.0364*** & $-$0.0166*** & $-$0.1859*** \\*
 & (0.0034) & (0.0060) & (0.0047) & (0.0053) \\
Wave 6 & $-$0.0812*** & $-$0.0598*** & $-$0.0174*** & $-$0.1664*** \\*
 & (0.0031) & (0.0055) & (0.0043) & (0.0049) \\
Wave 7 & $-$0.0785*** & $-$0.0245*** & $-$0.0405*** & $-$0.1707*** \\*
 & (0.0039) & (0.0069) & (0.0055) & (0.0062) \\
Wave 8 & $-$0.0580*** & $-$0.0389*** & 0.0297*** & $-$0.1648*** \\*
 & (0.0032) & (0.0055) & (0.0045) & (0.0051) \\
Wave 9 & $-$0.0247*** & $-$0.0124** & 0.0412*** & $-$0.1029*** \\*
 & (0.0032) & (0.0053) & (0.0044) & (0.0053) \\
Constant & 0.6682*** & 0.2735*** & 0.7905*** & 0.9407*** \\*
 & (0.0469) & (0.0751) & (0.0616) & (0.0744) \\
\addlinespace
Observations & 50{,}997 & 50{,}997 & 50{,}997 & 50{,}997 \\
Individuals & 21{,}605 & 21{,}605 & 21{,}605 & 21{,}605 \\
R-squared & 0.2760 & 0.2436 & 0.1715 & 0.1444 \\
\midrule
\multicolumn{5}{@{}l}{\textit{Panel B. East}} \\
CAPB (\% of GDP) & 0.0011** & 0.0031*** & 0.0031*** & $-$0.0030*** \\*
 & (0.0005) & (0.0010) & (0.0008) & (0.0008) \\
Female & 0.0377*** & 0.0153*** & 0.0389*** & 0.0587*** \\*
 & (0.0022) & (0.0038) & (0.0028) & (0.0032) \\
\textbf{CAPB $\times$ Female} & 0.0022*** & 0.0010 & 0.0016* & 0.0039*** \\*
 & (0.0006) & (0.0011) & (0.0008) & (0.0009) \\
Age & $-$0.0170*** & $-$0.0095*** & $-$0.0167*** & $-$0.0247*** \\*
 & (0.0013) & (0.0023) & (0.0018) & (0.0021) \\
Age squared (/100) & 0.0157*** & 0.0144*** & 0.0105*** & 0.0222*** \\*
 & (0.0009) & (0.0016) & (0.0013) & (0.0015) \\
Primary education & $-$0.0361*** & $-$0.0058 & $-$0.0379*** & $-$0.0648*** \\*
 & (0.0101) & (0.0177) & (0.0126) & (0.0167) \\
Secondary education & $-$0.1106*** & $-$0.0866*** & $-$0.1186*** & $-$0.1267*** \\*
 & (0.0093) & (0.0166) & (0.0116) & (0.0156) \\
Tertiary education & $-$0.1794*** & $-$0.1579*** & $-$0.2080*** & $-$0.1723*** \\*
 & (0.0095) & (0.0169) & (0.0118) & (0.0158) \\
Employed or self-employed & $-$0.0608*** & $-$0.1095*** & $-$0.0753*** & 0.0025 \\*
 & (0.0030) & (0.0053) & (0.0041) & (0.0044) \\
Other employment status & 0.1066*** & 0.1227*** & 0.1422*** & 0.0550*** \\*
 & (0.0038) & (0.0065) & (0.0052) & (0.0058) \\
Poland & 0.0654*** & 0.0385*** & 0.0932*** & 0.0645*** \\*
 & (0.0032) & (0.0055) & (0.0041) & (0.0049) \\
Slovenia & 0.0068** & $-$0.0401*** & 0.0701*** & $-$0.0095** \\*
 & (0.0029) & (0.0052) & (0.0037) & (0.0043) \\
Estonia & 0.0919*** & 0.1248*** & 0.1242*** & 0.0268*** \\*
 & (0.0029) & (0.0049) & (0.0036) & (0.0043) \\
Wave 6 & $-$0.0135*** & 0.0003 & $-$0.0376*** & $-$0.0032 \\*
 & (0.0017) & (0.0031) & (0.0026) & (0.0023) \\
Wave 7 & $-$0.0409*** & $-$0.0325*** & $-$0.0604*** & $-$0.0298*** \\*
 & (0.0040) & (0.0072) & (0.0056) & (0.0059) \\
Wave 8 & $-$0.0134*** & $-$0.0183*** & $-$0.0286*** & 0.0067** \\*
 & (0.0022) & (0.0039) & (0.0032) & (0.0032) \\
Wave 9 & $-$0.0124*** & $-$0.0335*** & $-$0.0415*** & 0.0379*** \\*
 & (0.0021) & (0.0037) & (0.0031) & (0.0032) \\
Constant & 0.7761*** & 0.4679*** & 0.9733*** & 0.8871*** \\*
 & (0.0486) & (0.0852) & (0.0664) & (0.0763) \\
\addlinespace
Observations & 59{,}472 & 59{,}472 & 59{,}472 & 59{,}472 \\
Individuals & 26{,}648 & 26{,}648 & 26{,}648 & 26{,}648 \\
R-squared & 0.2280 & 0.2039 & 0.1171 & 0.1050 \\
\midrule
\multicolumn{5}{@{}l}{\textit{Panel C. Continental/Nordic}} \\
CAPB (\% of GDP) & 0.0006 & 0.0001 & 0.0025*** & $-$0.0008 \\*
 & (0.0006) & (0.0012) & (0.0008) & (0.0010) \\
Female & 0.0293*** & 0.0172*** & 0.0259*** & 0.0448*** \\*
 & (0.0020) & (0.0035) & (0.0025) & (0.0030) \\
\textbf{CAPB $\times$ Female} & $-$0.0010 & 0.0005 & $-$0.0018* & $-$0.0016 \\*
 & (0.0008) & (0.0015) & (0.0010) & (0.0013) \\
Age & $-$0.0224*** & $-$0.0256*** & $-$0.0189*** & $-$0.0229*** \\*
 & (0.0011) & (0.0020) & (0.0013) & (0.0017) \\
Age squared (/100) & 0.0193*** & 0.0238*** & 0.0135*** & 0.0206*** \\*
 & (0.0008) & (0.0014) & (0.0009) & (0.0012) \\
Primary education & $-$0.0661*** & $-$0.0483*** & $-$0.0939*** & $-$0.0561*** \\*
 & (0.0076) & (0.0117) & (0.0096) & (0.0113) \\
Secondary education & $-$0.1115*** & $-$0.0830*** & $-$0.1529*** & $-$0.0985*** \\*
 & (0.0073) & (0.0112) & (0.0093) & (0.0108) \\
Tertiary education & $-$0.1621*** & $-$0.1472*** & $-$0.2119*** & $-$0.1273*** \\*
 & (0.0074) & (0.0114) & (0.0093) & (0.0110) \\
Employed or self-employed & $-$0.0551*** & $-$0.1036*** & $-$0.0632*** & 0.0014 \\*
 & (0.0026) & (0.0046) & (0.0032) & (0.0038) \\
Other employment status & 0.0831*** & 0.1057*** & 0.0983*** & 0.0454*** \\*
 & (0.0036) & (0.0062) & (0.0048) & (0.0050) \\
Sweden & $-$0.0473*** & $-$0.0979*** & $-$0.0252*** & $-$0.0187*** \\*
 & (0.0028) & (0.0048) & (0.0032) & (0.0040) \\
France & $-$0.0090*** & $-$0.0581*** & 0.0034 & 0.0277*** \\*
 & (0.0031) & (0.0054) & (0.0039) & (0.0045) \\
Denmark & $-$0.0319*** & $-$0.0705*** & $-$0.0298*** & 0.0046 \\*
 & (0.0029) & (0.0054) & (0.0034) & (0.0043) \\
Wave 5 & $-$0.0561*** & 0.0331*** & $-$0.0093*** & $-$0.1922*** \\*
 & (0.0021) & (0.0036) & (0.0028) & (0.0035) \\
Wave 6 & $-$0.0630*** & 0.0196*** & $-$0.0183*** & $-$0.1904*** \\*
 & (0.0022) & (0.0038) & (0.0029) & (0.0036) \\
Wave 7 & $-$0.0673*** & 0.0140** & $-$0.0289*** & $-$0.1871*** \\*
 & (0.0030) & (0.0055) & (0.0039) & (0.0046) \\
Wave 8 & $-$0.0531*** & 0.0326*** & $-$0.0073** & $-$0.1846*** \\*
 & (0.0026) & (0.0044) & (0.0033) & (0.0041) \\
Wave 9 & $-$0.0504*** & 0.0359*** & $-$0.0117*** & $-$0.1755*** \\*
 & (0.0027) & (0.0045) & (0.0034) & (0.0042) \\
Constant & 1.0129*** & 1.0769*** & 0.9767*** & 0.9852*** \\*
 & (0.0397) & (0.0701) & (0.0488) & (0.0597) \\
\addlinespace
Observations & 71{,}392 & 71{,}392 & 71{,}392 & 71{,}392 \\
Individuals & 29{,}652 & 29{,}652 & 29{,}652 & 29{,}652 \\
R-squared & 0.2277 & 0.1604 & 0.1087 & 0.1481 \\
\bottomrule
\multicolumn{5}{@{}p{\dimexpr\columnwidth-2\tabcolsep\relax}@{}}{\footnotesize\emph{Separate regressions by region, with full controls and country and wave fixed effects, so the country dummies differ across panels. South: Italy, Spain, Greece (Portugal dropped, Section 3.1). Reference categories: no education and retired in all panels; Spain (South), Czech Republic (East), Germany (Continental/Nordic); Wave 1 in South and Continental/Nordic, Wave 5 in East, where Wave 1 is not available. Standard errors in parentheses, clustered at the individual. * p\textless0.10, ** p\textless0.05, *** p\textless0.01.}}
\end{longtable}

\normalsize

\small
\begin{longtable}[]{@{}
  >{\raggedright\arraybackslash}p{(\columnwidth - 6\tabcolsep) * \real{0.4000}}
  >{\raggedright\arraybackslash}p{(\columnwidth - 6\tabcolsep) * \real{0.2000}}
  >{\raggedright\arraybackslash}p{(\columnwidth - 6\tabcolsep) * \real{0.2000}}
  >{\raggedright\arraybackslash}p{(\columnwidth - 6\tabcolsep) * \real{0.2000}}@{}}
\caption{Health sub-index, adding income position}\\
\toprule
\textbf{} & \textbf{Baseline} & \textbf{+ income} & \textbf{+ income int.} \\
\midrule
\endhead
CAPB (\% of GDP) & 0.0019*** & 0.0018*** & 0.0019*** \\*
 & (0.0005) & (0.0005) & (0.0005) \\
Female & 0.0199*** & 0.0188*** & 0.0187*** \\*
 & (0.0021) & (0.0021) & (0.0021) \\
\textbf{CAPB $\times$ Female} & 0.0011* & 0.0012** & 0.0013** \\*
 & (0.0006) & (0.0006) & (0.0006) \\
Bottom income quartile & -- & 0.0338*** & 0.0338*** \\*
 &  & (0.0021) & (0.0021) \\
CAPB $\times$ Bottom quartile & -- & -- & $-$0.0003 \\*
 &  &  & (0.0007) \\
Observations & 181,861 & 181,861 & 181,861 \\
R-squared & 0.2041 & 0.2057 & 0.2057 \\
\bottomrule
\multicolumn{4}{@{}p{\dimexpr\columnwidth-2\tabcolsep\relax}@{}}{\footnotesize\emph{Outcome: health sub-index. Income position is the bottom income quartile within country-wave. Standard errors in parentheses, clustered at the individual. Full controls and country and wave fixed effects. * p\textless0.10, ** p\textless0.05, *** p\textless0.01.}}
\end{longtable}

\normalsize

\footnotesize
\begin{longtable}[]{@{}
  >{\raggedright\arraybackslash}p{(\columnwidth - 12\tabcolsep) * \real{0.2600}}
  >{\raggedright\arraybackslash}p{(\columnwidth - 12\tabcolsep) * \real{0.1230}}
  >{\raggedright\arraybackslash}p{(\columnwidth - 12\tabcolsep) * \real{0.1230}}
  >{\raggedright\arraybackslash}p{(\columnwidth - 12\tabcolsep) * \real{0.1230}}
  >{\raggedright\arraybackslash}p{(\columnwidth - 12\tabcolsep) * \real{0.1230}}
  >{\raggedright\arraybackslash}p{(\columnwidth - 12\tabcolsep) * \real{0.1230}}
  >{\raggedright\arraybackslash}p{(\columnwidth - 12\tabcolsep) * \real{0.1250}}@{}}
\caption{Robustness of the pooled results}\\
\toprule
\textbf{} & \textbf{Level} & \textbf{Lag} & \textbf{Change} & \textbf{Pre-2020} & \textbf{No w9} & \textbf{Headc.} \\
\midrule
\endhead
CAPB (or variant) & 0.0027*** & 0.0026*** & 0.0006** & 0.0045*** & 0.0041*** & 0.0052*** \\*
 & (0.0003) & (0.0003) & (0.0003) & (0.0004) & (0.0003) & (0.0007) \\
\textbf{CAPB $\times$ Female (or variant)} & 0.0005 & 0.0003 & 0.0004 & $-$0.0000 & 0.0002 & 0.0001 \\*
 & (0.0003) & (0.0003) & (0.0004) & (0.0005) & (0.0004) & (0.0008) \\
Observations & 181,861 & 181,861 & 181,861 & 128,302 & 145,340 & 181,861 \\
R-squared & 0.2749 & 0.2747 & 0.2743 & 0.2783 & 0.2792 & 0.1911 \\
\bottomrule
\multicolumn{7}{@{}p{\dimexpr\columnwidth-2\tabcolsep\relax}@{}}{\footnotesize\emph{Level is the baseline specification. Lag and Change replace the CAPB level with its one-year lag and its year-on-year change, and the interaction row follows the same variant. Pre-2020 keeps interviews up to 2019. No w9 drops Wave 9. Outcome is the composite index except in Headcount, where it is the binary headcount (deprived in at least one third of weighted indicators). Standard errors in parentheses, clustered at the individual. * p\textless0.10, ** p\textless0.05, *** p\textless0.01.}}
\end{longtable}

\normalsize

\begin{center}
\begin{minipage}{\textwidth}
\captionof{table}{Sensitivity of inference to the clustering level}
\scriptsize
\setlength{\tabcolsep}{2.2pt}
\resizebox{\ifdim\width>\textwidth\textwidth\else\width\fi}{!}{%
\begin{tabular}{@{}lccccccccc@{}}
\toprule
 & & \multicolumn{2}{c}{Individual} & \multicolumn{2}{c}{Country-wave} & \multicolumn{2}{c}{Country} & \multicolumn{2}{c}{Bootstrap (Webb)} \\
\cmidrule(lr){3-4}\cmidrule(lr){5-6}\cmidrule(lr){7-8}\cmidrule(lr){9-10}
Estimate & Coef. & SE & $p$ & SE & $p$ & SE & $p$ & $p$ & 95\% CI \\
\midrule
\multicolumn{10}{@{}l}{\textit{Panel A. Pooled sample}} \\
CAPB, composite & 0.0027 & 0.0003 & $<$0.0001 & 0.0009 & 0.0045 & 0.0010 & 0.0213 & -- & -- \\
CAPB $\times$ Female, composite & 0.0005 & 0.0003 & 0.1651 & 0.0005 & 0.3780 & 0.0006 & 0.4675 & 0.4388 & [$-$0.0014, 0.0017] \\
CAPB $\times$ Female, health & 0.0011 & 0.0006 & 0.0555 & 0.0008 & 0.1506 & 0.0011 & 0.3116 & 0.3866 & [$-$0.0011, 0.0039] \\
CAPB $\times$ Female, economic & $-$0.0021 & 0.0004 & $<$0.0001 & 0.0009 & 0.0286 & 0.0008 & 0.0204 & 0.0115 & [$-$0.0036, $-$0.0005] \\
CAPB $\times$ Female, social & 0.0024 & 0.0005 & $<$0.0001 & 0.0016 & 0.1345 & 0.0025 & 0.3675 & 0.5253 & [$-$0.0041, 0.0079] \\
\midrule
\multicolumn{10}{@{}l}{\textit{Panel B. By welfare-regime group, separate regressions}} \\
CAPB $\times$ Female, South, health & 0.0005 & 0.0008 & 0.5328 & 0.0014 & 0.7230 & -- & -- & 0.7421 & -- \\
CAPB $\times$ Female, South, economic & 0.0009 & 0.0006 & 0.1844 & 0.0011 & 0.4447 & -- & -- & 0.4745 & -- \\
CAPB $\times$ Female, South, social & 0.0053 & 0.0008 & $<$0.0001 & 0.0023 & 0.0344 & -- & -- & 0.1026 & -- \\
CAPB $\times$ Female, East, health & 0.0010 & 0.0011 & 0.3625 & 0.0014 & 0.4772 & -- & -- & 0.4972 & -- \\
CAPB $\times$ Female, East, economic & 0.0016 & 0.0008 & 0.0525 & 0.0014 & 0.2684 & -- & -- & 0.3084 & -- \\
CAPB $\times$ Female, East, social & 0.0039 & 0.0009 & $<$0.0001 & 0.0033 & 0.2529 & -- & -- & 0.2581 & -- \\
CAPB $\times$ Female, Continental/Nordic, health & 0.0005 & 0.0015 & 0.7077 & 0.0019 & 0.7720 & -- & -- & 0.7850 & -- \\
CAPB $\times$ Female, Continental/Nordic, economic & $-$0.0018 & 0.0010 & 0.0766 & 0.0011 & 0.1022 & -- & -- & 0.1055 & -- \\
CAPB $\times$ Female, Continental/Nordic, social & $-$0.0016 & 0.0013 & 0.1984 & 0.0018 & 0.3703 & -- & -- & 0.4074 & -- \\
\bottomrule
\addlinespace[3pt]
\multicolumn{10}{@{}p{\dimexpr\columnwidth-2\tabcolsep\relax}@{}}{\footnotesize\emph{Point estimates are identical across clustering levels; only the variance estimator changes. Clusters: 77,905 individuals, 58 country-waves, 11 countries. Panel A bootstrap: wild cluster bootstrap by country, Webb weights, 9,999 replications, seed 20260927; the $p$-value and confidence set refer to the interaction. With 11 countries the country-clustered standard errors are under-clustered, so read them alongside the bootstrap. Panel B bootstrap: wild cluster bootstrap on country-wave, Webb weights (17 country-waves in South and East, 24 in Continental/Nordic). Country-level clustering is not reported by region, since each region has only three or four countries.}} \\
\end{tabular}}
\end{minipage}
\end{center}

\footnotesize
\begin{longtable}[]{@{}
  >{\raggedright\arraybackslash}p{(\columnwidth - 4\tabcolsep) * \real{0.5000}}
  >{\raggedright\arraybackslash}p{(\columnwidth - 4\tabcolsep) * \real{0.2500}}
  >{\raggedright\arraybackslash}p{(\columnwidth - 4\tabcolsep) * \real{0.2500}}@{}}

\caption{Sensitivity to Slovenia's group assignment}\\
\toprule
\textbf{} & \textbf{With Slovenia} & \textbf{Without Slovenia} \\
\midrule
\endhead
Pooled sample, composite & 0.0005 & 0.0005 \\*
 & (0.0003) & (0.0004) \\
East, composite & 0.0022*** & 0.0033*** \\*
 & (0.0006) & (0.0010) \\
East, health & 0.0010 & 0.0006 \\*
 & (0.0011) & (0.0018) \\
East, economic & 0.0016* & 0.0015 \\*
 & (0.0008) & (0.0014) \\
East, social & 0.0039*** & 0.0077*** \\*
 & (0.0009) & (0.0016) \\
\midrule
Observations, pooled sample & 181{,}861 & 167{,}861 \\
Observations, East & 59{,}472 & 45{,}472 \\
\bottomrule
\multicolumn{3}{@{}p{\dimexpr\columnwidth-2\tabcolsep\relax}@{}}{\footnotesize\emph{Each cell is the $CAPB \times Female$ coefficient, with its standard error below, from a separate regression with full controls and country and wave fixed effects. Without Slovenia drops Slovenia from the sample. Standard errors are clustered at the individual. * p\textless0.10, ** p\textless0.05, *** p\textless0.01.}}
\end{longtable}

\normalsize